\documentclass[12pt]{article}

\newcommand {\be}{\begin{equation}}
\newcommand {\ee}{\end{equation}}
\newcommand{\bea}{\begin{eqnarray}}
\newcommand{\eea}{\end{eqnarray}}
\newcommand{\ba}{\begin{array}}
\newcommand{\ea}{\end{array}}

\renewcommand{\c}{{\rm c}}
\newcommand{\const}{{\rm const}}
\newcommand{\n}{{\bf n}}

\begin{document}
\noindent {\Large\bf The Calculation of Critical Parameters in
SU(2) Gauge Theory with Kouvel-Fisher Method}

\vspace{0.3cm} \noindent {O.A. Mogilevsky}

\vspace{0.3cm}

\noindent {\it Bogolyubov  Institute for Theoretical Physics, 14-B
Metrologichna Str., Kiev 03143, Ukraine}

\begin{abstract}
We calculate the critical coupling $4/g_c^2$ and critical
exponent $\beta$ for the order parameter in SU(2) lattice gauge
theory by applying of the finite size scaling technique and
the method proposed by Kouvel and Fisher for analysis of
experimental data. In contrast to the standard finite size
scaling approach, this method allows to determine simultaneously
both $g_c^2/4$ and $\beta$ as two parameters of the linear fit to
the Monte-Carlo data.
\end{abstract}

Lattice calculations by Monte-Carlo (MC) method have so far been
the only method for studying the temperature phase transition in
gauge theories from the first principles. By their nature, MC
calculations are carried out on the finite lattices. This is a
most serious drawback of a computer simulations, because no
finite system can exhibit a true phase transition. Nevertheless,
finite systems remind of phase transitions and systematic studies
of these pseudo-transitions as function of system size may reveal
information about the phase transition in the thermodynamic
limit. One way to do this particularly in order to evaluate the
critical exponents at the theory is to use finite size scaling
(FSS).

FSS was proposed in statistical physics [1, 2] and applied to
high precision MC data of Ising [3, 4] and other models. The
validity of this method for gauge theories has been demonstrated
through the investigation of critical properties of SU(2) lattice
gauge theories at finite temperature [5--9].

In this letter we propose the method of analysis of MC data which
is the combination of FSS technique and so-called Kouvel-Fisher
method for analysis of experimental data in solid state physics
[11]. We consider MC data for SU(2) order parameter on the
lattices $8^3\times4$, $12^3\times4$ and $18^3\times4$. In
contrast to the standard FSS approach, our method allows to
determine the critical coupling $4/g_\c^2$ and critical exponent
$\beta$ as  two parameters of the linear fit.

The partition function for SU(2) gauge theory on
$N_\sigma^3\times N_\tau$ lattice is
\be
Z=\int\prod\limits_{(\mu,\nu)} d U_{\mu,\nu} e^{-S(U)},
\ee
where $U_{\mu,\nu}$ are the SU(2) matrices on the link
$(\mu,\nu)$, $S$ is the standard Wilson action
\be
S(U)={4\over g^2}\sum_p \left(1-{1\over2} Tr Up \right).
\ee
$Up$ is a product of link matrices around a plaquette. The
temperature $T$ (volume $V$) is defined by size of the lattice in
the time-like (space-like) direction ($a$ is a lattice spacing)
\be
T=1/\left(N_\tau a\right), \qquad V=\left(N_\sigma a\right)^3.
\ee
The order parameter of the deconfinement phase transition on an
infinite volume lattice is the expectation value of lattice
average of the Polyakov loop
\be
\langle L\rangle = \Biggl\langle {1\over N_\sigma^3} \sum_{\n}
L_\n\Biggr\rangle,
\ee
where
\be
L_\n={1\over2}Tr \prod\limits_{\tau=1}^{N_\tau} U_{\tau,\n;0}
\ee
and $U_{\tau,\n;0}$ is the SU(2) link matrix at the point
$(\n,\tau)$ in the time-like direction. On the finite lattice one
considers as the order parameter $\langle|L|\rangle$, since in
this case $\langle L\rangle$ is always zero due to system flips
between the two ordered states.

According to FSS theory the scaling function for the order
parameter on the lattice $N_\sigma^3\times N_\tau$, $N_\tau$ fix
is given by [2]
\be
N_\sigma^{{\beta\over\nu}}\langle|L|\rangle =Q_L\left(
x,g_1N_\sigma^{-y_1}\right),
\ee
where $\beta$, $\nu$ are the critical exponents for order
parameter and correlation length correspondingly, $g_1$ is the
irrelevant scaling field, $y_1\approx1$. The scaling variable $x$
is defined as
\be
x=t\cdot N_\sigma^{{1\over\nu}}, \qquad t={g_\c^2/4 -g^2/4\over
g_\c^2/4}\, .
\ee
Our definition of reduce temperature $t$ differs from one used in
[7--10]
\be
t={4/g^2 -4/g_\c^2\over 4/g_\c^2}
\ee
and seems to be more natural because the coupling constant
$g^2/4$ plays just the role of a "temperature" of the effective
theory which can be obtained from the partition function (1) by
integration on all link variables except those for Polyakov
loops. It is clear that this difference disappears in the limit
$g^2/4\rightarrow g_\c^2/4$ .

At large $N_\sigma$ the scaling function $Q_L$ must behave as
\be
Q_L\sim (x)^\beta\, .
\ee
This behaviour may be observed only if $x$ becomes "sufficiently
large". Deviations from (9) are in close vicinity of the critical
point due to finite size rounding and further away from $x=0$ due
to correction to scaling. Expanding $Q_L$ in (6) around $x=0$ at
large $N_\sigma$ results in
\be
N_\sigma^{{\beta\over\nu}}\langle|L|\rangle =a+{b\over
N_\sigma^{y_1}}+ \left(c+{d\over N_\sigma^{y_1}}\right)x,
\ee
where $a$, $b$, $c$, $d$ are unknown constants. Dropping the term
$O(N_\sigma^{-y_1})$ one can obtain
\be
N_\sigma^{{\beta\over\nu}}\langle|L|\rangle =a+cx,
\ee
what is exactly the approach of Refs. [7--10]. On the other hand
at $x$ "sufficiently large" the $x$-dependence is drastically
changed. In Ref. [6] the scaling function $Q_L$ has been
calculated in the form
\be
Q_L=A x^\beta\left[ 1+{B\over N_\sigma^{y_1}}x^{y_1\nu}\right]
\ee
by the data collapsing method. It has been shown that the
correction to scaling term in (12) is negligibly small as
compared to the leading term on the interval $0,5<x<2$. In both
cases the SU(2) Monte-Carlo data for the order parameter have been
used which were computed on $N_\sigma^3\times N_\tau$ lattices
with $N_\sigma$=8,12,18,26 and $N_\tau=4$ and have already
reported on in Refs. [5--10]. The numerical value of the critical
exponent $\beta$ has been found which is compatible with the
corresponding value $\beta=0,327$ of the 3-$d$ Ising model [4].

These two procedures have two disadvantages: i) $3-d$ Ising model
value of $\nu=0,631$ has to be used; ii) the very accurate
calculation of critical coupling $g_\c^2/4$ is necessary. The
latter is determining by the fourth cumulant of $\langle|L|\rangle$
as in Refs. [5, 7] or by the $\chi^2$-method as proposed in Ref.
[8]. Even small shift leads to the additional uncertainty in the
determination of critical exponent $\beta$, because it depends
very sensitively on the value inserted for $g_\c^2/4$.

We propose below the method for the determination of the critical
exponent $\beta$ which does not require to know of $g_\c^2/4$
from the very beginning. No information about the critical
exponent $\nu$ is also required. Following the Kouvel-Fisher idea
which was used for analysis of the experimental data in solid
state physics [11] we consider the function
\be
K(x)={Q_L\over \partial Q_L/\partial x}\, .
\ee
According to the results of Ref. [6] in the region $0,5<x<2$ it
is easy to find from (12)
\be
K(x)={x\over\beta} \left[1+{\const\over
N_\sigma^{y_1}} x^{y_1\nu}\right].
\ee
Dropping the correction to scaling term we have in this region
with the same accuracy as in (11)
\be
K(x)={x\over\beta}.
\ee
Taking into account the identity
\be
\frac{\partial Q_L}{\partial x}=-{g_\c^2\over4}
N_\sigma^{-{(1-\beta)\over\nu}}\  \frac{\partial
\langle|L|\rangle}{\partial (g^2/4)}
\ee
we have immediately
\be
K(x)={x\over\beta} =-{N_\sigma^{{1\over\nu}}\over g_\c^2/4}
\ {\langle|L|\rangle\over \partial
\langle|L|\rangle/\partial (g^2/4)}.
\ee
Finally, we obtain the simple but informative expression
\be
{\langle|L|\rangle\over \partial
\langle|L|\rangle/\partial (g^2/4)} ={1\over\beta}
\left({g^2\over4}-{g_\c^2\over4}\right).
\ee
Eq. (18) means that for the lattices with $N_\tau$ fix and
various $N_\sigma$ there are the intervals of $g^2/4$ where MC
data for different $N_\sigma$ must arrange around the same
straight line. The inverse slop of this line represents the
critical exponent $\beta$ and the intersection with $g^2/4$ axis
yields the critical coupling.

In order to verify this approach we have used the high precision
MC data produced by Bielefeld group on the lattices $8^3\times4$,
$12^3\times4$ and $18^3\times4$ [5, 10]. The calculation of
derivatives with respect to $g^2/4$ has been made by using the
parabolic fit for $\langle|L|\rangle$ (every parabolae includes
five MC points). This fit does not smooth the statistical
fluctuations, but connects the data continuously and with
continuous first and second derivatives. The least minimal
$\chi^2$-fit of Eq. (18) ($\chi^2/N=0,62$) includes 7 MC points in
the region $0.396<g^2/4<0.426$ ($2.345<4/g^2 <2.527$) for
$N_\sigma=8$, 24 MC points in the region $0.414<g^2/4<0.430$
($2.325<4/g^2<2.415$) for $N_\sigma=12$ and 16 MC points in the
region $0.427<g^2/4<0.433$ ($2.310<4/g^2<2.340$) for
$N_\sigma=18$. The result is
\be
\beta=0.326\pm0.003; \qquad 4/g_\c^2=2.2988\pm 0.0007.
\ee
We see that the critical exponent $\beta$ is in excellent
agreement with $3-d$ Ising model value and the inverse critical
coupling $4/g^2_\c$ coincides with that value which has been
calculated in Refs. [7--10] by other method.

In conclusion the following comments are in order. We have shown
that the Kouvel-Fisher modification of FSS gives the extremely
useful method for the calculation of critical exponent $\beta$ in
SU(2) lattice gauge theory. This method allows to investigate the
finite lattice system at small but finite
$t=(g^2/4-g_\c^2/4)/(g_\c^2/4)$ and large $N_\sigma$ with
infinite volume formulae (18). The critical coupling $g_\c^2/4$
and critical exponent $\beta$ are simultaneously determined as
two parameters of the linear fit. It is clear that this approach
may be applied to the calculation of all critical exponents in
SU(2) gauge theory and to other models which undergo the
second-order phase transition.

\end{document}